\documentclass[aip,reprint, jmp, amsmath, amssymb]{revtex4-1}
\usepackage{graphicx}% Include figure files
\usepackage{dcolumn}% Align table columns on decimal point
\usepackage{amsmath}
\usepackage{bm}% bold math
\usepackage{txfonts}
\begin{document}

\title{Realization of polarization evolution on higher-order Poincar\'{e} sphere with metasurface}% Force line breaks with \\

\author{Yachao Liu}
\author{Xiaohui Ling}
\author{Xunong Yi}
\author{Xinxing Zhou}
\author{Hailu Luo}
\email{hailuluo@hnu.edu.cn}
\author{Shuangchun Wen}
\email{scwen@hnu.edu.cn} \affiliation{Key Laboratory for
Micro-/Nano-Optoelectronic Devices of Ministry of Education, College
of Physics and Microelectronic Science, Hunan University, Changsha
410082, China}
\date{\today}% It is always \today, today,

\begin{abstract}
We present a simple and convenient method to yield cylindrical
vector (CV) beams and realize its polarization evolution on
higher-order Poincar\'{e} sphere based on inhomogeneous birefringent
metasurface. By means of local polarization transformation of the
metasurface, it is possible to convert a light beam with homogeneous
elliptical polarization into a vector beam with any desired
polarization distribution. The Stokes parameters of the output light
are measured to verify our scheme, which show well agreement with
the theoretical prediction. Our method may provide a convenient way
to generate CV beams, which is expected to have potential
applications in encoding information and quantum computation.
\end{abstract}

\maketitle

Cylindrical vector (CV) beams are light beams of which the
polarization states are arranged with cylindrical symmetry in the
beam cross section~\cite{Zhan2009}. A plenty of unique properties
originated from their special intrinsic symmetry have distinguished
the CV beams from general optical beams with homogeneous
polarization. For example, the radially polarized light, a subset of
the CV beams, can lead to a strong longitudinal field in the case of
strong focusing~\cite{Novotny2001,Dorn2003}. Because of these
particularities, CV beams are expected with a broad applications
such as particle trapping~\cite{Zhan2004}, high resolution
imaging~\cite{Novotny2001}, particle acceleration~\cite{Varin2002},
and microscopy~\cite{Abouraddy2006}.

Similar to the geometric representation of homogeneous polarizations
on Poincar\'{e} sphere, a prominent geometric representation of the
CV beams is provided by the so-called higher-order Poincar\'{e}
sphere (HOPS)~\cite{Holleczek2011,Milione2011}. In this geometrical
representation rule, higher-order Pancharatnam-Berry phase is
demonstrated by cyclic transformation of the CV beams on the
HOPS~\cite{Berry1984,Milione2012}. This phase is geometric in nature
and differs significantly from a dynamic phase, which is
proportional to light's total angular momentum. This state evolution
will provide an advantage solution to create optical qubits, a
single photon carrying several bits of information, which is
expected to improve the flexibility of information encoding and
simplify quantum computation~\cite{Leek2007}. Nevertheless, there is
still not an effective method to generate all the states on HOPS and
realize the state evolution. Most efforts including using
subwavelength nanostructure~\cite{Bomzon2001,Beresna2011,Iwami2012},
orientation-tailored liquid crystal
~\cite{Marrucci2006,Chen2011,Ling2012},
interferometry~\cite{Wang2007,Ruiz2013}, laser intracavity
devices~\cite{Oron2000}, and fiber laser~\cite{Fridman2008,Zhou2009}
have been made to obtain the radial and azimuthal polarized beams,
however, the other states are seldom referred to.

In this work, we demonstrate a simple and convenient method to
generate the CV beams and realize the evolution of polarization
states on the HOPS. It is well known that the cascaded polarizer,
quarter-wave plate (QWP), and half-wave plate (HWP) can generate any
elliptical polarization state at will~\cite{Damask2005}.
Furthermore, by modulating the direction of the optical axis of the
QWP/HWP, the polarization can evolve along the longitude/latitude on
the surface of the fundamental Poincar\'{e} sphere. Analogously,
while it comes to the CV beams whose polarization distribution are
location-related, a special waveplate with space-variant optical
axes should be employed. Fortunately, as a two dimensional
electromagnetic nanostructure, metasurface with tailorable structure
geometry posseses unparalleled advantages in optical
phase~\cite{Ni2013,Huang2013} and polarization~\cite{Chen2012}
manipulation, especially in subwavelength
scale~\cite{Yu2011,Kildishev2013,Shitrit2013}, and is particularly
amenable to produce and manipulate the CV beams.

\begin{figure}
\includegraphics[width=8.5cm]{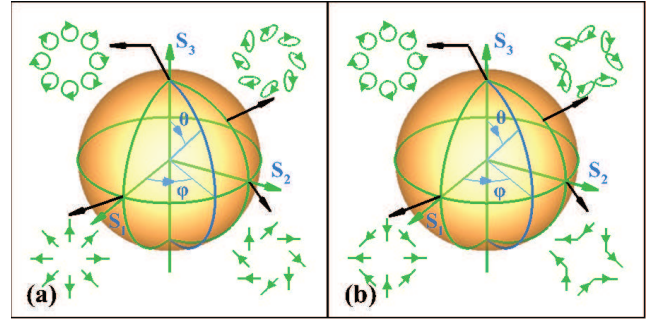}
\caption{\label{Fig1} Schematic illustration of the HOPS with (a)
$l=+1$ and (b) $l=-1$, respectively. The polarization distribution
of four different points on each sphere are shown, and ($\theta$,
$\varphi$) are the spherical coordinates of HOPS.}
\end{figure}

We first consider the polarization transformation of a HWP with its
fast and slow optical axes lying parallel to the $xy$ coordinate
plane. The Jones matrix of a HWP can be written as
\begin{equation}
\left(\begin{array}{ccc}
\cos{2\Phi} & \sin{2\Phi} \\
\sin{2\Phi} & -\cos{2\Phi}
\end{array} \right),
\end{equation}
where $\Phi$ is the orientation of the fast optical axis. Without
loss of generality, a homogeneous elliptical polarization light is
taken into account. It can be represented geometrically on the
fundamental Poincar\'{e} sphere and algebraically described by the
following equation in terms of the azimuthal and polar angles
($\vartheta$, $\alpha$) in the sphere,
\begin{equation}
|{\psi}({\vartheta},{\alpha})\rangle=\cos{\left(\frac{\vartheta}{2}\right)}|{R}\rangle{e}^{\textit{i}\sigma\frac{\alpha}{2}}
+\sin{\left(\frac{\vartheta}{2}\right)}|{L}\rangle{e}^{-\textit{i}\sigma\frac{\alpha}{2}}\label{1}.
\end{equation}
Here $|{R}\rangle=(\hat{x}+\textit{i}\hat{y})/\sqrt{2}$ and
$|{L}\rangle=(\hat{x}-\textit{i}\hat{y})/\sqrt{2}$ are respectively
the right and left circular polarizations. The constant
$\sigma=\pm1$ is associated with the spin angular momentum with
$\sigma\hbar$ per photon. When this beam impinges on the HWP at
normal incidence, the resulting polarization is still elliptical
with its polarization given by
\begin{equation}
|{\psi}({\vartheta'},{\alpha'})\rangle=\cos{\left(\frac{\vartheta'}{2}\right)}|{R}\rangle{e}^{-\textit{i}(\sigma\frac{\alpha'}{2}+2\Phi)}
+\sin{\left(\frac{\vartheta'}{2}\right)}|{L}\rangle{e}^{\textit{i}(\sigma\frac{\alpha'}{2}+2\Phi)}\label{2},
\end{equation}
where $\vartheta'=\pi-\vartheta$, $\alpha'=-\alpha$. It also
represents an elliptical polarization located on the point
($\vartheta', \alpha'$) of the fundamental Poincar\'{e} sphere.
Actually, all the left-handed photons transform into right-handed
photons and acquire an additional phase $\exp(\pm2\Phi)$, and vice
versa. As a result, the magnitude of the two components have also
exchanged.

\begin{figure}
\includegraphics[width=8.5cm]{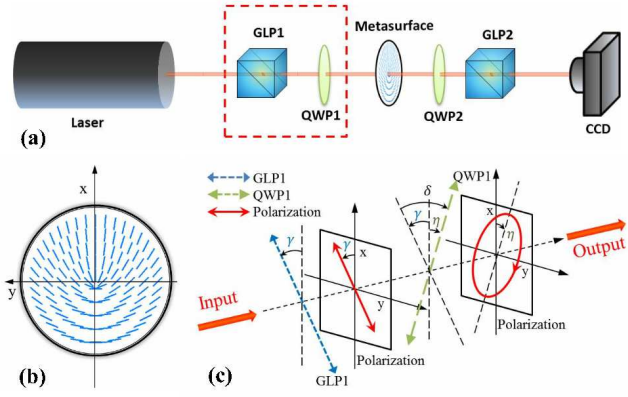}
\caption{\label{Fig2} (a) The experiment setup to generate CV beams.
The He-Ne laser produces a linearly polarized light at 632.8nm
wavelength (17 mW, Thorlabs HNL210L-EC). GLP, Glan laser polarizer;
HWP, half waveplate; QWP, quarter waveplate; CCD, charge-coupled
device (Coherent LaserCam HR). (b) Schematic drawing of the optical
axes in metasurface. (c) The angular dependence of GLP, QWP, and the
generated elliptical polarization.}
\end{figure}

More interestingly, providing the HWP exhibits a space-variant
optical axis direction, e.g., $\Phi=q\phi$ with $\phi=\arctan(y/x)$
the local azimuthal angle and $q$ is an integer or a semi-integer,
the phase factor $\exp(\pm2\Phi)$ indicates an optical vortex with
its topological charge $l=\pm2q$. As the two orthogonal circular
polarization components carry just opposite vortex phase, their
coherent superposition actually forms a space-variant, inhomogeneous
polarization beam, i.e., CV beam. In this case, Eq.~(\ref{2})
represents an arbitrary elliptical CV beam on the HOPS (see
Fig.~\ref{Fig1}). This involves three free parameters $\vartheta'$,
$\alpha'$, and $\Phi$, which serve as effective ways to manipulate
the CV beams. The first two parameters are related to the incident
polarization which can be controlled by a polarizer and a QWP [see
Fig.~\ref{Fig2}(a)]. As $\Phi$ determines the topological charge,
with different $\Phi$, the CV beams located on the HOPS with
different topological charges $l$. Once the inhomogeneous HWP is
manufactured, we can obtain any desired CV beam with a fixed
topological charge and realize the polarization evolution on the
HOPS.

We now can derive what kind of incident homogenous polarization on
the fundamental Poincar\'{e} sphere should be given, when a CV beam
located on the point ($\vartheta', \alpha'$) of the HOPS is desired.
Comparing the Eq.~(\ref{2}) to (\ref{1}), we can easily obtain the
relationship: $\vartheta=\pi-\vartheta'$ and $\alpha=-\alpha'$.
However, in most cases, it is more effective to characterize the
elliptical polarization with azimuth angle $\eta$ and elliptical
angle $\delta$. According to our calculation, the $\eta$ and
$\delta$ of incident light can be associated with the spherical
coordinates ($\vartheta$, $\alpha$) of HOPS in the forms,
\begin{equation}
{\eta=\alpha'=-\alpha},
\end{equation}
\begin{equation}
{\delta=-\frac{1}{2}(\vartheta-\frac{\pi}{2})}.
\end{equation}

We implement an experiment to demonstrate this idea. The
experimental apparatus is shown in Fig.~\ref{Fig2}(a). First, an
elliptical polarization beam is produced by a polarizer (GLP1) and a
quarter-wave plate (QWP1), and then passes through a inhomogeneous
HWP. The inhomogeneous HWP now can be conveniently realized by an
artificial inhomogeneous metasurface which is fabricated by etching
space-variant grooves on a fused silica sample using a femtosecond
laser~\cite{Beresna2011}. This artificially creates an inhomogeneous
form birefringence on the isotropic sample, and the local direction
of the optical axes (slow and fast axes) are perpendicular and
parallel to the grooves, respectively. This dielectric-based
metasurface makes up a converter which can generate a CV beam when
an elliptically polarized light is illuminated. To verify the
viability of our scheme, we select a metasurface with $q=0.5$ and
homogeneous $\pi$ retardation (Altechna). Figure~\ref{Fig2}(b) shows
the schematic drawing of the local optical axes. The Stokes
parameters of the resulting beam is measured by a typical setup
(QWP2, GLP2, and CCD).

By simply rotating the GLP1 or the QWP1, the azimuth angle or the
ellipticity of the output elliptical polarization will change
continuously, and we can get any elliptical polarization states from
the first part of the setup, which is summarily illustrated in the
Fig.~\ref{Fig2}(c). The azimuth angle $\eta$ of the output
polarization ellipse is the same as the optical axis direction of
the QWP1 and the ellipticity angle $\delta$ equals to the relative
optical axis direction of the GLP1 and the QWP1. At the same time,
the final state output from the metasurface will evolve on HOPS
along the latitude or the longitude. In this way, we can generate
arbitrary states on HOPS and realized the polarization state
evolution on the HOPS.

\begin{figure}
\includegraphics[width=8.5cm]{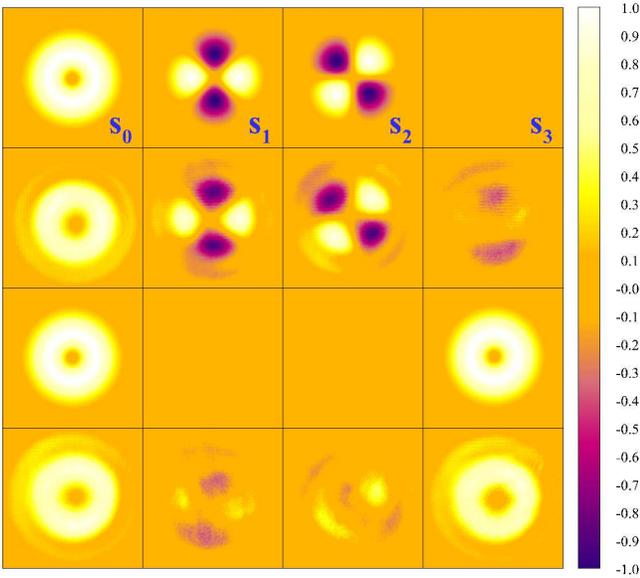}
\caption{\label{Fig3} Normalized Stokes parameters $s_{0}$, $s_{1}$,
$s_{2}$, and $s_{3}$. The first and second rows are theoretical and
experimental results of point the (1,~0,~0), respectively. The third
and forth rows are the theoretical and experimental results
corresponding to the point (0,~0,~1). Both points are on the surface
of the HOPS ($l=+1$).}
\end{figure}

\begin{figure}[t]
\includegraphics[width=8.5cm]{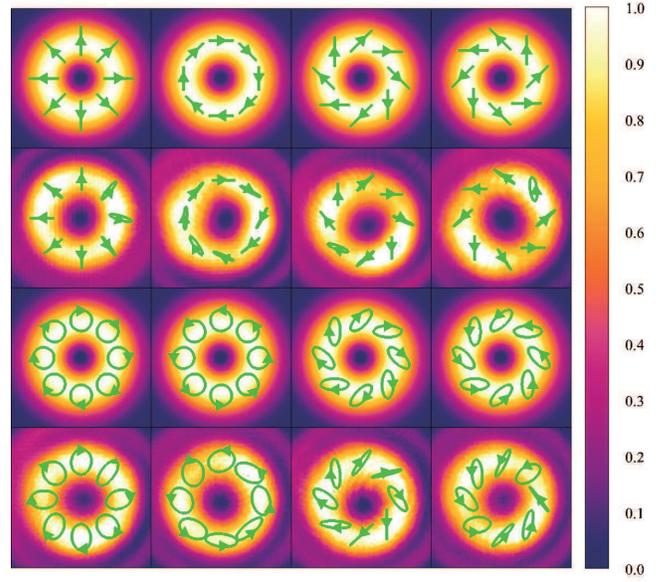}
% Here is how to import EPS art
\caption{\label{Fig4} The polarization and intensity distribution of
the theoretical and experimental results. The first and second rows
are respectively the theoretical and experimental results of the
points $(1,~0,~0)$, $(-1,~0,~0)$, $(0,~1,~0)$, and $(0,~-1,~0)$ in
the order from left to right. The third and forth rows are the
corresponding theoretical and experimental results of points
$(0,~0,~1)$, $(0,~0,~-1)$, $(0,~\sqrt{2}/{2},~\sqrt{2}/{2})$, and
$(0,~\sqrt{2}/{2},~-\sqrt{2}/{2})$. All the points are on the
surface of the HOPS ($l=+1$).}
\end{figure}

To effectively analyze the emerging polarization states, we measured
the Stokes parameters. They are a set of indices ($S_0$, $S_1$,
$S_2$, and $S_3$) that describe the polarization of light. When the
optical axes of the QWP2 and polarizer (GLP2) make up the following
combinations, what we get in the CCD camera are exactly the
parameters~\cite{Born1997},
\begin{equation}
S_{1}=I(0^\circ,0^\circ)-I(90^\circ,90^\circ),
\end{equation}
\begin{equation}
S_{2}=I(45^\circ,45^\circ)-I(135^\circ,135^\circ),
\end{equation}
\begin{equation}
S_{3}=I(-45^\circ,0^\circ)-I(45^\circ,0^\circ),
\end{equation}
where $I(\alpha,\beta)$ represents the intensity what we get when
the optical axis of QWP2 is $\alpha$ with the fixed $x$ axis and the
polarization direction of GLP2 is $\beta$ with respect to the same
$x$ axis. The first Stokes parameter $S_{0}$ is the intensity
distribution of the output beam which can be recorded by the CCD
without QWP2 and GLP2.

Figure~\ref{Fig3} shows the normalized Stokes parameters
$s_i=S_i/S_0$ ($i=1,2,3$) of two points on the HOPS ($l=+1$). It is
known to all that $s_1$ describes the linear polarization in $x$
($s_i=+1$) or $y$ ($s_i=-1$) direction, $s_2$ linear polarization in
$\pm45^{\circ}$ ($s_2=\pm1$) respect to the $x$ direction, and $s_3$
the degree of the circular polarization with $s_3=\pm1$
corresponding to left- and right-handed polarization, respectively.
Point (1,~0,~0) on the HOPS represents the radially polarization
with axial symmetry [see Fig.~\ref{Fig1}(a)]. The $s_1$ and $s_2$
show a four-lobe pattern, with alternative $+s_{1,2}$ and $-s_{1,2}$
components. In this case $s_3$ equals to 0, because the light beam
is essential a linear polarization. While for point (0,~0,~1) which
locates on the north pole of the HOPS, represents the circularly
polarized vortex, so the $s_{1,2}=0$ and the $s_3$ equals to the
total intensity of the light ($s_0$). In comparison with the
theoretical distribution, the experimental results evidenced that
the emerging beams are the desirable CV beams. For further
verification, we mapped the polarization distribution. Firstly, we
extracted the exact parameter values of any states from the Stokes
parameters. Then, according to the relationship of the Stokes
parameters and the polarization, the polarization states of the
emerging beams are worked out. By depicting the graph of
polarization distribution, we intuitively verify that the output
beams are the desired CV beams. Figure~\ref{Fig4} theoretically and
experimentally record the polarization states of eight points on the
HOPS ($l=+1$). These points compose of two special paths on the
sphere, one is on the equator and the other is on the longitude of
azimuth angle $\varphi=45^{\circ}$. Comparatively, the experimental
results agree well with the theory. The deviation of the
experimental results is because that, when extracting the parameter
values, it is difficult to ensure the pictures located at the same
pixel.

Although our experiments are limited to realizing the state
evolution on HOPS with $l=+1$, our scheme is able to produce the
states on HOPS with $l=-1$ [Fig.~\ref{Fig1}(b)] and other larger
$|l|$. It is well known that HWP would inverse the chirality of the
passing light~\cite{Damask2005}. By inserting a HWP after the
metasurface in our setup, we can change the sign of the topological
charge. To generate vector beams with larger topological charge
$|l|$, a metasurface with a larger $q$ is needed. In general, it is
theoretically possible to produce any states on all HOPSs, by
inserting a HWP and a metasurface with suitable $q$ values in our
setup.

In conclusion, we have presented a simple and flexible method to
produce arbitrary states on HOPS by using a birefringent
metasurface. Within our setup, corresponding elliptical polarization
can be converted to the desirable CV beam. The experimental results
show well agreement with the theoretical ones, which is proved by
measuring the Stokes parameters of the tested states. Furthermore,
by simply rotating a GLP or QWP in our setup, the produced state
will evolve along the latitude or longitude of the HOPS. It is
believed that our scheme shall have potential applications in
designing vector beams with more complex polarization distribution,
encoding information, and quantum computation.

This research was partially supported by the National Natural
Science Foundation of China (Grants Nos. 61025024, 11274106 and
11347120), and the Scientific Research Fund of Hunan Provincial
Education Department of China (Grant No. 13B003).

\end{document}